# Energy Spectrum of Local Multiparticle Configurations and the Mechanism of Anomalously Slow Relaxation of the System of Strongly Interacting Liquid Clusters in a Disordered Nanoporous Medium According to the Self-Organized Criticality Scenario

V. D. Borman, V. N. Tronin, V. A. Byrkin

*Department of Molecular Physics, National Research Nuclear University MEPhI, Kashirskoe sh. 31, Moscow 115409, Russia*

**Abstract**

It has been shown that changes in the energy of a system of nonwetting-liquid clusters confined in a random nanoporous medium in the process of relaxation can be written in the quasiparticle approximation in the form of the sum of the energies of local (metastable) configurations of liquid clusters interacting with clusters in the connected nearest pores. The energy spectrum and density of states of the local configuration have been calculated.

It has been shown that the relaxation of the state of the system occurs through the scenario of self-organized criticality (SOC). The process is characterized by the expectation of a fluctuation necessary for overcoming a local energy barrier of the metastable state with the subsequent rapid hydrodynamic extrusion of the liquid under the action of the surface buoyancy forces of the nonwetting framework. In this case, the dependence of the interaction between local configurations on the number of filled pores belonging to the infinite percolation cluster of filled pores serves as an internal feedback initiating the SOC process. The calculations give a power-law time dependence of the relative volume $\theta$ of the confined liquid $\theta \sim t^{-\alpha}$ ($\alpha \sim 0.1$).

The developed model of the relaxation of the porous medium with the nonwetting liquid demonstrates possible mechanisms and scenarios of SOC for disordered atomic systems.

## 1. Introduction

Anomalously slow relaxation is a common property of many disordered condensed media with various sizes of particles from atomic to micron. These media include metallic and molecular glasses, polymers, colloid media, granulated matter, nanocomposites, etc. The relaxation of these media is characterized by the «stretched exponential law»; i.e., a characteristic property depends on the time as $\exp(-(t/\tau_0)^\beta)$, where $\tau$ is the characteristic relaxation time and $\beta < 1$ (see, e.g., [1, 2]). For this reason, anomalously slow relaxation is considered as a cooperative phenomenon [3]. Several phenomenological models have been proposed to describe anomalously slow relaxation. They were analyzed in several recent reviews [2–5]. The key assumption in these models is the assumption of the existence of local configurations of particles that are not detected in experiments on the measurement of the static density correlation function.

In this work, we study a disordered nanoporous medium (silica gel) formed at a sol–gel transition in a colloid medium. The random structure of pores and framework of the medium «is frozen» at gelling owing to a chemical reaction between colloid particles. A liquid that fills the space of pores if pores form a connected system can inherit the random structure. This occurs at a porosity above the percolation threshold, when the infinite percolation cluster of connected pores is formed in the medium and, thereby, the filling of pores of the medium with the liquid is ensured. The characteristics of the structure of pores such as the specific volume of pores, specific surface of pores, and pore size distribution are studied by the methods of gas adsorption–desorption [6, 7] and liquid porometry [8–10]. The nonwetting liquid (water) filling the porous medium at an increased pressure is used in liquid porometry. In this application, the liquid is dispersed and forms an ensemble of interacting clusters. When excess pressure is reduced to zero, the liquid should be extruded from pores under the action of the surface buoyancy forces of the hydrophobic framework of the medium. New information on the disordered structure can be obtained by studying the confinement of the nonwetting liquid in the random structure of pores [11–14]. According to [13, 14], this phenomenon is due to the effective interaction (attraction) between liquid clusters in neighboring pores, which is determined by correlation in the arrangement of pores. An anomalously slow relaxation of the state of the confined liquid was observed in [15] as the slow extrusion of the liquid from the porous medium with the inverse power-law dependence of the volume of the confined liquid on the time with the exponent less than 0.1. Depending on the properties of the porous medium, nonwetting liquids can remain in it for many days and even months [16–19]. A small change in the temperature or a decrease in the degree of filling near the critical value results in the rapid extrusion of the liquid (dispersion transition) [11]. In this work, analyzing the dynamics of a change in the energy spectrum of the states of the set of degenerate local metastable configurations of interacting liquid clusters in neighboring pores, we propose a mechanism of this phenomenon. These local configurations interact with each other through the field of the infinite percolation cluster of filled pores according to the arrangement of pores in the frozen random structure of pores in the porous medium. Since the liquid in the frozen random structure of pores follows this structure, it could be expected that the mechanisms of anomalously slow relaxation of the framework in the processes of gelling and the confinement of the liquid in pores should be the same.

One of the main results of the numerical experiments on studying the mechanism of anomalous relaxation of random media is the space–time heterogeneity of such media (dynamic heterogeneity (DH) and dynamic facilitation (DF) models [2–4, 20–22]). This space–time local heterogeneity follows from the statistical model named mode-coupling theory (MCT) [2, 4, 5, 21, 22]. Another property of MCT is the divergence of the characteristic relaxation time $\tau$ at the critical temperature predicted in MCT. This divergence is attributed in MCT to blocking of the motion of particles by neighboring particles because of memory effects at their interaction [5]. However, this is observed neither in experiments nor in numerical calculations [4].

In the phenomenological model called random first-order transition (RFOT) theory (a first-order phase transition in a random medium) [1–5, 23, 24]) of relaxation in the disordered medium, the system of particles with a short-range interaction between them is considered as a system of local regions in different metastable states. The RFOT model was formulated under the assumption of the possible realization of the state of glass with a short-range interaction between particles as a heterogenic state with numerous local configurations similar to the disordered state of spin glass in the Potts model with the *p*-spin of particles and their long-range (infinite-range) interaction [5]. Because of thermal fluctuations in the RFOT model, the states of these local regions of glass change, they can transform to each other, and large regions can decay into small regions. Thus, a hypothetical «mosaic» structure of the state of disordered glass appears with local regions whose size increases with a decrease in the temperature. It is assumed that the time scale of a transition between the states of local regions coincides with the scale of the characteristic relaxation time depending on the temperature.

This model has not yet been statistically justified, and some numerical studies [25] within the achieved computation time does not confirm the mosaic structure of the state. The calculation of the energy of local metastable configurations would be of a key importance for the confirmation of the RFOT model. This model predicts the existence of the temperature at which the divergence of the characteristic relaxation time should be observed, but is not observed in fact.

Various variants of local regions in the disordered medium were revealed and studied in numerical experiments [20, 26, 27]. The relation between geometrical frustrations and static local structures with the characteristic relaxation time was analyzed in [28]. It was found that geometrical frustrations could be responsible for a change in the scenario of cooperative relaxation. In view of the results of that work, the authors assumed that anomalously slow relaxation could possibly have a noncooperative character.

The above brief review demonstrates that the nature of local heterogenic regions, the reason for their appearance and metastability, and the nature and mechanism of anomalously slow relaxation in disordered glass media remain unclear. It is also unclear whether the experimentally observed properties are equilibrium or kinetic in nature.

Another approach to the description of anomalously slow relaxation of the confined liquid is possible within the phenomenological theory of self-organized criticality (SOC) [29, 30]. This model was proposed in [31, 32] to describe relaxation as an avalanche process of falling of a pile of sand. Self-organized criticality was later used to describe the relaxation in nonlinear nonequilibrium systems in a critical state. It is commonly accepted [33, 34] that the SOC state appears in nonlinear dissipative systems in the critical state that relax without the external control parameter through rapid avalanche transitions between different metastable states of a system. The critical state is maintained not at a point, but in a wide region of the phase diagram of states through a nonlinear feedback mechanism because of the existence of the internal mechanism leading to the dynamic self-organization of transitions between intermediate metastable states of the system. Self-organized criticality is characteristic of the systems with fractal objects. However, physical mechanisms responsible for avalanche transitions between metastable states, as well as the reason for the appearance of the feedback mechanism leading to the dynamic self-organization of transitions between such states, are not yet understood. In particular, numerical studies [35] indicate that the properties of SOC with avalanche relaxation for spin-glass systems are not manifested at a finite number of neighbors and a divergent number of neighboring interacting spins are necessary.

In this work, we propose a new mechanism of anomalous relaxation of states of a liquid confined in a random nanoporous medium, which can be characterized as the SOC mechanism in revealed properties. It is shown that the state of the nonwetting liquid confined in a random system of connected nanopores is the state of strongly interacting liquid clusters in pores connected through the percolation cluster, which should exist for the filling of all pores of the porous medium and extrusion from them. The interaction energy is calculated as the energy of the total surface of liquid clusters in neighboring pores. Change in the energy of such a system can be represented in the form of the sum of the energies of individual local «excitations», i.e., local metastable configurations of a liquid cluster in a nanopore and its environment. Since the structure of pores in the porous medium is random and the infinite cluster of filled pores is fractal, such local configurations differ from each other in radii of pores, number of the nearest neighbor pores, and number of filled and empty pores in them. This consideration is close to a usual quasiparticle approach in the problem of strongly interacting particles [36]. The energy spectrum of these configurations and the density of states are calculated in the low polydispersity approximation for the case of a narrow pore size distribution with the mean radius $\bar{R}$ and a small relative width $\Delta R / \bar{R} < 0.1$.

The relaxation of the state of the nonwetting liquid confined in the random system of pores is manifested as a change in the volume of the liquid with the time $\theta(t)$. It was shown in [15] that the relaxation of the nonwetting liquid confined in the porous medium takes several stages. At the first stage, the nonwetting liquid flows out of weakly bound states of filled pores with a decrease in the pressure in the hydrodynamic time $\tau_0 \sim 0.1$ s[12]. This stage ends with the formation of the critical metastable state of filled pores in the porous medium, which slowly relaxes in time $t_0 \gg \tau_0$ and, then, decays at times $t > t_0$.

The relaxation of the state under consideration at times $t > t_0$ is due to the decay of local metastable configurations and was described within the formalism of extrusion time distribution functions [38]. The process of decay involves waiting for thermal fluctuation necessary for overcoming the metastable state and the subsequent rapid hydrodynamic extrusion of the liquid, which is a dissipative avalanche transition to a new metastable state. The decay of local metastable configurations is accompanied by a decrease in the number of filled pores in the percolation cluster and, thereby, in the number of neighboring pores. As a result, the energy of the multiparticle interaction (attraction) between liquid clusters in neighboring pores decreases. Consequently, the decay of certain local metastable configurations successively creates critical conditions for the decay of other local metastable configurations with a different, initially higher energy. Such a relaxation of the state of the nonwetting liquid confined in the random system of pores at times much larger than the time of the hydrodynamic extrusion of the liquid from pores occurs according to the SOC mechanism. The dependence of the interaction between local configurations on the number of filled pores belonging to the infinite percolation cluster of filled pores serves as an internal feedback initiating the SOC process.

The calculations give a power-law time dependence of the relative volume $\theta$ of the confined liquid $\theta \sim t^{-\alpha}$ ($\alpha \sim 0.1$). Such a dependence is characteristic of known self-organized criticality phenomena.

## 2. The quasiparticulate approach to calculation of energy of liquid in the disordered confinement of a random porous medium

We assume that the disordered nanoporous medium includes pores with different sizes and the size of the porous medium is much larger than the maximum size of pores, so that the porous medium can be considered as infinite and the infinite percolation cluster of empty pores is formed in it.

The time of the extrusion of the liquid from a filled pore in such a medium can be calculated within the statistical theory of fluctuations. According to [43], the probability $w$ of a change in the state of the system in unit time at the extrusion of the liquid from the pore under consideration because of fluctuations in the system is determined by a change in the entropy of the system at the extrusion of the liquid from the pore $w \sim exp(\Delta S)$. The proportionality coefficient in this relation is determined by the dynamics of extrusion of the liquid. According to a small thermal effect observed in the experiments on the extrusion of the liquid [37], the probability can be represented in the form $w = w_0 exp(-\delta A/T)$, where $\delta A$ is the isothermal work that should be spent on the extrusion of the liquid from the pore. The quantity inverse to the probability $w$ of extrusion of the liquid from pores in unit time determines the extrusion time $\tau = \tau_0 exp(\delta A/T)$. The hydrodynamic extrusion time of the liquid can be estimated by the formula $\tau_0 = \dfrac{4\pi R^3}{Q(R)}$, where $Q(R) = \dfrac{\pi}{8\eta_0}\dfrac{\Delta p}{L} R^4$ is the rate of the dissipative flow of the liquid with the viscosity $\eta$ in the channel with the radius $R$, length $L$, and pressure drop $\Delta p$ [38]. According to this formula, $\tau_0 \sim 1/R$ and, at $\Delta p \sim \sigma \cos\vartheta/R$ for water with $\sigma = 72 mJ/m^2$ [42] and $\vartheta = 110^0$ [11], $\tau_0 \sim 0.1 нсек$ [12, 13].

The work $\delta A$ serves as a potential barrier for the extrusion of the liquid from the pore. If $\delta A < 0$, the characteristic extrusion time is determined by the hydrodynamic time of the motion of the liquid in the porous medium, $\tau \sim \tau_0$. If $\delta A > 0$, the characteristic liquid extrusion time is determined by the potential barrier for extrusion $\delta A > 0$. The characteristic liquid extrusion time in this case can be much larger than the hydrodynamic time, $\tau >> \tau_0$, when the barrier is above the temperature $\delta A > T$. The barrier $\delta A$ includes the work ($pV$) done by the system on an increase in its volume by the volume $V$ of the pore at the pressure $p$ and a change in the surface energy $\delta\varepsilon$ of the liquid in the pore at the extrusion of the liquid. The quantity $\delta A$ at zero pressure is equal to the change in the energy $\delta\varepsilon$ at the extrusion of the liquid from the pore. In this case, the state of pores in the nearest environment of the depleted pore changes. We will show that the change in the energy $\delta\varepsilon$ of the porous media with the nonwetting liquid at the extrusion of the liquid from filled pores is represented in the form of the sum of the energies of local metastable configurations.

According to [12, 13], the metastable state of filled pores in the porous medium is formed at times $t > \tau_0$, relaxes slowly at times much larger than $\tau_0$, and then decays at times $t > t_0$. It was shown in [12] that the relaxation of the metastable state in the Libersorb23–water system with a narrow pore size distribution with the relative width $\Delta R/\bar{R} < 0.1$ occurs at times $t_0 \sim 10^5$ s much larger than the thermal equilibrium establishment time $\tau_e < \tau_0$ at the extrusion of the liquid from the pore $t_0 \gg \tau_0, \tau_e$. Therefore, such a state of the disordered porous medium with the confined nonwetting liquid can be described by local equilibrium distribution functions and can be considered thermodynamically. We consider the spatially disordered porous medium consisting of $N$ pores with various random radii $R_i$ and porosity $\varphi$ filled to the degree of filling $\theta$. Each pore with the radius $R_i$ located at the point $\vec{r}_i$ can be either filled or empty. We introduce the number $n_i$ that is unity if this pore is filled with the liquid and zero if it is empty. The phase space of such a system is $5N$-dimensional space of the coordinates of the pores, their radii, and filling factors $n_i$ of all $N$ pores. Let $F(\vec{r}_1, R_1, \vec{r}_2, R_2 \ldots \vec{r}_N, R_N, n_1, n_2 \ldots n_N)$ be the $N$-particle distribution function of pores in their coordinates $\vec{r}_i$, radii $R_i$, and filling factors $n_i$. The function $F(\vec{r}_1, R_1, \vec{r}_2, R_2 \ldots \vec{r}_N, R_N, 0_1, 0_2 \ldots 0_N)$ is the $N$-particle distribution function of empty pores and $F(\vec{r}_1, R_1, \vec{r}_2, R_2 \ldots \vec{r}_N, R_N, 1_1, 1_2 \ldots 1_N)$ is the distribution function of pores in the completely filled porous medium. At $\theta < 1$, the state of the porous medium can correspond to different geometrical configurations of filled and empty pores. Hence, the multiparticle distribution function of empty and filled pores $F_\theta(\vec{r}_1, R_1, \vec{r}_2, R_2 \ldots \vec{r}_N, R_N)$ for the porous medium filled to the degree of filling $\theta$ is degenerate and can be obtained from $F(\vec{r}_1, R_1, \vec{r}_2, R_2 \ldots \vec{r}_N, R_N, n_1, n_2 \ldots n_N)$ by summing over such configurations $F_\theta(\vec{r}_1, R_1, \vec{r}_2, R_2 \ldots \vec{r}_N, R_N) = \sum_{k=1}^{N_\theta} F(\vec{r}_1, R_1 \ldots \vec{r}_N, R_N \{n_i\}_\theta^k)$. The distribution function $F(\vec{r}_1, R_1 \ldots \vec{r}_N, R_N \{n_i\}_\theta^k)$ corresponds to $\{n_i\}_\theta^k$ configurations of empty and filled pores such that the total relative volume of filled pores is $\theta$, the index $k = 1 \ldots N_\theta$ marks such a configuration, and $N_\theta$ is the maximum number of degenerate configurations. In particular, in the case of nonoverlapping pores, the multiparticle distribution function of empty and filled pores $F_\theta(\vec{r}_1, R_1, \vec{r}_2, R_2 \ldots \vec{r}_N, R_N)$ for the porous medium with the degree of filling $\theta$ has the form

$$F_\theta(\vec{r}_1, R_1, \vec{r}_2, R_2 \ldots \vec{r}_N, R_N) = \sum_{\{n_i\}} F(\vec{r}_1, R_1, \vec{r}_2, R_2 \ldots \vec{r}_N, R_N, \{n_i\}) \delta\left(\frac{\sum_{i=1}^{N} n_i V_i}{\sum_{i=1}^{N} V_i} - \theta\right),$$ where $V_i$ is the

volume of the $i$ th pore. For the filling of pores in the porous medium, it should contain an infinite cluster through which the transport of the liquid is possible [39, 40]. Consequently, the percolation cluster of filled pores is formed inside this percolation cluster of pores at filling [12]. According to the form of the multiparticle distribution function of empty and filled pores $F_\theta(\vec{r}_1, R_1, \vec{r}_2, R_2 \ldots \vec{r}_N, R_N)$ in the porous medium with the degree of filling $\theta$, such a state can be degenerate because different realizations of states of the percolation cluster can correspond to the degree of filling $\theta$.

The complete thermodynamic potential, i.e., the energy of the porous medium filled to the degree of filling $\theta$, can be written in the form

$$E = \int d\Gamma \sum_{k=1}^{N_\theta} \varepsilon(\vec{r}_1, R_1, \vec{r}_2, R_2 \ldots \vec{r}_N, R_N \{n_i\}_\theta^k) F_\theta(\vec{r}_1, R_1, \vec{r}_2, R_2 \ldots \vec{r}_N, R_N \{n_i\}_\theta^k). \tag{1}$$

We assume that the extrusion of the liquid from a randomly chosen filled pore with the radius $R_1$ located at the point $\vec{r}_1$ in the porous medium does not change the state of the other $N-1$ pores. In this case, the distribution function $F_\theta(\vec{r}_1, R_1, \vec{r}_2, R_2 \ldots \vec{r}_N, R_N)$ is represented in the form of the product of the single-particle distribution function $f(r_1, R_1)$ corresponding to the depleted pore and ($N-1$)-particle function $F_\theta(\vec{r}_2, R_2 \ldots \vec{r}_N, R_N)$ describing the state of the remaining medium. According to [13], a change in the energy $\delta E$ of the system can then be written in the form

$$\delta E = \int \delta\varepsilon(\vec{r}_1, R_1) f(\vec{r}_1, R_1) d\vec{r}_1 dR_1,$$
$$F_\theta(\vec{r}_1, R_1, \vec{r}_2, R_2 \ldots \vec{r}_N, R_N) = f(r_1, R_1) F_\theta(\vec{r}_2, R_2 \ldots \vec{r}_N, R_N). \quad (2)$$

Here, $f(\vec{r}_1, R_1)$ is the single-particle distribution function of filled pores normalized to the total number of filled pores in the porous medium filled to the degree of filling $\theta$ and $\delta\varepsilon(\vec{r}_1, R_1)$ is the change in the energy of the system after the depletion of one pore,

$$\delta\varepsilon(\vec{r}_1, R_1, \theta) = \int d\vec{r}_2 \ldots d\vec{r}_{N-1} dR_2 \ldots dR_{N-1} \varepsilon(\vec{r}_1, R_1, \vec{r}_2, R_2 \ldots \vec{r}_N, R_N, \theta)$$
$$\delta\varepsilon(\vec{r}_1, R_1 \ldots \vec{r}_N, R_N, \theta) = \sum_{k=1}^{N_\theta} \varepsilon(\vec{r}_1, R_1, \vec{r}_2, R_2 \ldots \vec{r}_N, R_N \{\Delta n_1 = 1, n_2 \ldots n_{N-1}\}_\theta^k) \times \quad (3)$$
$$\times F_\theta(\vec{r}_2, R_2 \ldots \vec{r}_{N-1}, R_{N-1} \{n_2 \ldots n_{N-1}\}_\theta^k)$$

Here, $\delta\varepsilon(\vec{r}_1, R_1 \ldots \vec{r}_N, R_N, \theta)$ is the sum of the energy $\delta\varepsilon(\vec{r}_1, R_1 \ldots \vec{r}_N, R_N, \theta)$ of the interface between the porous medium and liquid and the energy $\delta\varepsilon_{int}(\vec{r}_1, R_1 \ldots \vec{r}_N, R_N, \theta)$ necessary for the formation of menisci in the throats of the neighboring pores.

In this case, it is assumed that the chemical potential of the liquid does not change at its dispersion. This is valid for pores with the sizes $R > 1$ nm [17]. The energies $\delta\varepsilon_1(\vec{r}_1, R_1 \ldots \vec{r}_N, R_N)$ and $\delta\varepsilon_{int}(\vec{r}_1, R_1 \ldots \vec{r}_N, R_N)$ can be written in the form

$$\delta\varepsilon(\vec{r}_1, R_1 \ldots \vec{r}_N, R_N) = -\delta\sigma(1 - \eta(\vec{r}_1, R_1 \ldots \vec{r}_N, R_N)) S(\vec{r}_1, R_1), \quad \eta = \frac{S_m(\vec{r}_1, R_1 \ldots \vec{r}_N, R_N)}{S(\vec{r}_1, R_1)}, \quad (4)$$
$$\delta\varepsilon_{int}(\vec{r}_1, R_1 \ldots \vec{r}_N, R_N) = \sigma \delta S_m(\vec{r}_1, R_1 \ldots \vec{r}_N, R_N)$$

Here, $\sigma$ is the surface energy of the liquid, $\delta\sigma = (\sigma_{ls} - \sigma_{sg})$ is the difference between the surface energies of the solid–liquid and solid–gas interfaces, $S(\vec{r}_1, R_1)$ is the area of the surface of depleted pore, and $S_m(\vec{r}_1, R_1 \ldots \vec{r}_N, R_N)$ and $\delta S_m(\vec{r}_1, R_1 \ldots \vec{r}_N, R_N)$ are the area of menisci in the pore and the change in the area of menisci at the depletion of the pore, respectively. We assume that the area of menisci and the change in the area of menisci at depletion of the pore depends only on the nearest environment of the depleted pore. In this case, the $N$-particle distribution function $F_\theta(\vec{r}_1, R_1 \ldots \vec{r}_N, R_N)$ can be represented in the form of the product of the $z$-particle function $F_z(\vec{r}_1, R_1 \ldots \vec{r}_z, R_z)$ and ($N-z$)-particle function $F_{N-z}(\vec{r}_{z+2}, R_{z+2} \ldots \vec{r}_N, R_N)$, where $z$ is the number of pores in the environment of the depleted pore. Effects associated with the flow of the liquid to pores next of the nearest environment of the depleted pore without leaving the porous medium are neglected. In this case, $\delta\varepsilon(\vec{r}_1, R_1)$ can be represented in the form

$$\delta\varepsilon(\vec{r}_1,R_1) = -\delta\sigma(1-\langle\eta(\vec{r}_1,R_1)\rangle)S(\vec{r}_1,R_1) + \delta\varepsilon_{int}.$$

$$\langle\eta(\vec{r}_1,R_1)\rangle = \frac{\langle S_m(\vec{r}_1,R_1\ldots\vec{r}_N,R_N)\rangle}{S(\vec{r}_1,R_1)}. \qquad (5)$$

$$\delta\varepsilon_{int}(\vec{r}_1,R_1) = \sigma\langle\delta S_m(\vec{r}_1,R_1\ldots\vec{r}_N,R_N)\rangle = \sigma\langle W(\vec{r}_1,R_1\ldots\vec{r}_N,R_N)S_m(\vec{r}_1,R_1\ldots\vec{r}_N,R_N)\rangle$$

Here, averaging is performed with the distribution function $F_\theta(\vec{r}_1,R_1\ldots\vec{r}_N,R_N) = F_z(\vec{r}_1,R_1\ldots\vec{r}_z,R_z)F_{N-z}(\vec{r}_{z+2},R_{z+2}\ldots\vec{r}_N,R_N)$ and $W(\vec{r}_1,R_1\ldots\vec{r}_N,R_N)$ is the change in the number of menisci at the extrusion of the liquid from the filled pore.

To calculate such average values, we assume that the area $S_m$ of menisci in the environment of the pore with the radius $R_1$ is the sum of the areas of the menisci formed by mutually independent pores with the radius $R_2$ overlapping with the pore of the radius $R_1$. In this case, the distribution function $F_z(\vec{r}_1,R_1\ldots\vec{r}_z,R_z)$ can be represented in the form of the product of pair correlation functions $g_2(\vec{r}_1,R_1,\vec{x},R_2) = \int d\vec{r}_3 dR_3\ldots\vec{r}_N dR_N F_\theta(\vec{r}_1 R_1\ldots\vec{r}_N R_N)$ corresponding to correlations of the pore with the radius $R_1$ located at the point $\vec{r}_1$ and the pore with the radius $R_i, i = 2\ldots z$, located at the point $\vec{r}_i$:
$F_z(\vec{r}_1,R_1\ldots\vec{r}_z,R_z) = g_2(\vec{r}_1,R_1,\vec{r}_2,R_2)F_{z-1}(\vec{r}_3,R_3..\vec{r}_z,R_z) + ..g_2(\vec{r}_1,R_1,\vec{r}_z,R_z)F_{z-1}(\vec{r}_2,R_2,\vec{r}_3,R_3..\vec{r}_{z-1},R_{z-1})$. This means that correlations of the pore with the radius $R_1$ and each of the $z$ nearest mutually uncorrelated pores in its environment are taken into account, whereas triple and higher-order correlations of pores with each other are neglected. In this case, the average area of menisci $<S_m>$ can be written in the form

$$\langle S_m(\vec{r}_1,R_1\ldots\vec{r}_N,R_N)\rangle =$$
$$= \int F_\theta(\vec{r}_1,R_1\ldots\vec{r}_N,R_N)\sum_{k=1}^{z} s_m(\vec{r}_1,R_1,\vec{r}_k,R_k)d\vec{r}_2,dR_2\ldots d\vec{r}_N,dR_N = \qquad (6)$$

$$= \int d\vec{x}dR_1 s_m(\vec{r}_1,R_1,\vec{x},R_2)g_2(\vec{r}_1,R_1,\vec{x},R_2)$$

Here, $s_m(\vec{r}_1,R_1,\vec{x},R_2)$ is the area of the meniscus formed by the pores with the radii $R_1,R_2$ located at the points $\vec{r}_1$ and $\vec{r}_2$, respectively; $z$ is the number of the pores in the environment of the depleted pore; $g_2(\vec{r}_1,R_1,\vec{r}_2,R_2)$ is the pair correlation function of pores in the porous medium filled to the degree of filling $\theta$; and integration is performed over distances corresponding to the first coordination sphere $|\vec{x}| < R_1 + R_2$. For the spatially isotropic medium, $g_2(\vec{r}_1,R_1,\vec{r}_2,R_2) = g_2(R_1,|\vec{r}_2-\vec{r}_1|,R_2)$.

Within the model of randomly distributed spheres, the pair correlation function of the pores with the radii $R$ and $R_1$ and centers spaced by the distance $\vec{r}$ has the form [40, 41]

$$g_2(R,R_1,\vec{r}) = \varphi^{\frac{1}{R_1^3}\left(R^3+R_1^3-3/4x^2(R_1-x/3)-3/4y^2(R-y/3)\right)}, \quad x = \frac{R^2-(|\vec{r}|-R_1)^2}{2|\vec{r}|}, \quad y = R+R_1-x-|\vec{r}|. \qquad (7)$$

The integration of pair distribution function (7) over the volume nearest to the depleted pore gives the number of pores with the radius $R_1$ that are the nearest neighbors of the pore with the radius $R$

$$z(R, R_1) = \frac{1}{\varphi V_{pore}} \int_{|R-R_1|}^{|R+R_1|} g_2(R, R_1, \vec{r}) \ldots d\vec{r} \,. \tag{8}$$

Here, $V_{pore}$ is the volume of one pore with the radius $R_1$ and $\varphi$ is the porosity.

The pore with the radius $R$ can be surrounded by pores with various radii. For this reason, the average number of the nearest neighbor pores of the pore with the radius $R$ can be obtained by averaging Eq. (8) with the pore size distribution function $f(R_1)$ normalized to unity:

$$z(R) = \int_0^\infty dR_1 f(R_1) z(R, R_1) \,. \tag{9}$$

Expression (9) for the average number of the nearest neighbors $z(R)$ was analyed in [42].

We assume that the liquid is removed from a pore only if at least one of its neighboring pores belongs to the infinite cluster of filled pores through which the extrusion of the liquid from a granule of the porous medium is possible. In this case, a change in the energy of the pore because of a change in the number of menisci, $\sigma \langle \delta S_m(\vec{r}_1, R_1 \ldots \vec{r}_N, R_N) \rangle = \sigma \langle W(\vec{r}_1, R_1 \ldots \vec{r}_N, R_N) S_m(\vec{r}_1, R_1 \ldots \vec{r}_N, R_N) \rangle$, is nonzero only for the states that include at least one realization of the infinite cluster of filled pores. These states are determined by the distribution function $F_\theta(\vec{r}_1, R_1, \vec{r}_2, R_2 \ldots \vec{r}_N, R_N)$ for the pores belonging to the percolation cluster through which the liquid is extruded. The state of the system containing the infinite cluster of filled pores can have many realizations; i.e., this state is degenerate. Consequently, this state can be characterized by the probability $P(\theta)$ for the pore to belong to the infinite cluster of filled pores. The distribution function $F_\theta(\vec{r}_1, R_1 \ldots \vec{r}_N, R_N)$ for these states is represented in the form of the product of the $z$-particle function $F_z$ and $(N - z)$-particle function $F_{N-z}$, each being a functional of the probability $P(\theta)$ for the pore to belong to the infinite cluster of filled pores $F_z = F_z[P(\theta)], F_{N-z} = F_{N-z}[P(\theta)]$. The functional $F_z = F_z[P(\theta)]$ for $F_z(\vec{r}_1, R_1 \ldots \vec{r}_z, R_z) = g_2(\vec{r}_1, R_1, \vec{r}_2, R_2) g_2(\vec{r}_1, R_1, \vec{r}_3, R_3) .. g_2(\vec{r}_1, R_1, \vec{r}_z, R_z)$ includes all powers of the probability $P(\theta)$ from zero to $z$. When calculating averages entering into Eq. (5), we assume that $\langle W(\vec{r}_1, R_1 \ldots \vec{r}_N, R_N) S_m(\vec{r}_1, R_1 \ldots \vec{r}_N, R_N) \rangle \approx \langle W(\vec{r}_1, R_1 \ldots \vec{r}_N, R_N) \rangle \langle S_m(\vec{r}_1, R_1 \ldots \vec{r}_N, R_N) \rangle$. Then, using the representation of the function $F_z$ in the form of the product of the pair correlation functions, we conclude that $W = \langle W(\vec{r}_1, R_1 \ldots \vec{r}_N, R_N) \rangle$ is the functional of the probability $P(\theta)$ depending on the number $z$ of the nearest neighbors of the depleted pore with the radius $R$:

$$W = W(z(R), P(\theta)) \eta(R) \,. \tag{10}$$

Here, $W(z, \theta_1)$ is the difference between the average numbers of menisci before and after the depletion of the pore per nearest neighbor. The product of $W(z, \theta_1)$ by the surface energy of the liquid in menisci determines a change in the energy of the pore at the extrusion of the liquid from it. This quantity can be treated as the energy of the interaction $\delta \varepsilon_{int}$ of the liquid cluster in the pore with a liquid cluster in neighboring pores. According to Eqs. (5) and (6), in the approximations under consideration,

$$\begin{aligned} \delta \varepsilon(R, \theta_1) &= \delta \varepsilon_1(R) + \delta \varepsilon_{int}(R, \theta_1) \\ \delta \varepsilon_1(R) &= -\delta \sigma (1 - \eta(R)) S, \quad \eta(R) = \frac{\langle S_m(R, R_1) \rangle}{S} \\ \delta \varepsilon_{int}(R, \theta_1) &= \sigma \langle W S_m \rangle \approx \sigma W(z, \theta_1) \eta(R) \end{aligned} \tag{11}$$

Here, the area of menisci $S_m$ is determined by the nearest environment of the given pore. Therefore,

$$\eta(R) = \frac{1}{4\pi R^2}\int_0^\infty z(R,R_1) s_m(R,R_1) f(R_1) dR_1, \qquad (12)$$

where $s_m(R,R_1)$ is the area of one meniscus in the throat of the given pore with the radius $R$ connected with a pore of the radius $R_1$. Thus, to calculate the energy $\delta\varepsilon(R,\theta_1)$, it is necessary to calculate the change in the number of menisci at the depletion of the pore $W(z,\theta_1)$ and the connectivity factor $\eta(R)$. The quantity $W(z,\theta_1)$ was calculated in [12] as

$$W = \sum_{n=0}^{z} (1-\theta)^n (P(\theta))^{z-n} \frac{z-2n}{z} \frac{z!}{n!(z-n)!}. \qquad (13)$$

Here, $n$ is the number of neighboring empty pores, which can vary from zero to $z$. It follows from Eq. (13) that Eqs. (5) specify the energy of the pore taking into account multiparticle effects of the interaction between empty and filled pores in its environment. The probability $P(\theta)$ for the pore to belong to the infinite cluster of filled pores is a monotonically increasing function of the degree of filling $\theta$ such that $P(\theta) = 1$ at $\theta = 1$.

According to Eq. (2), a change in the energy of the porous medium filled to the degree of filling $\theta$ at the extrusion of the liquid from $k$ pores under the condition that the degree of filling $\theta$ changes only slightly can be written in the form

$$\delta E = \int \delta\tilde{\varepsilon}(R_1....R_k,\vec{r}_1..\vec{r}_k,\theta) F(R_1..R_k,\vec{r}_1..\vec{r}_k) dR_1 d\vec{r}_1..dR_k d\vec{r}_k$$

$$\delta\tilde{\varepsilon}(R_1....R_k,\vec{r}_1..\vec{r}_k) = \sum_{i=1}^{k} \delta\varepsilon(R_i,\vec{r}_i,\theta), \quad F(R_1..R_k,\vec{r}_1..\vec{r}) = \prod_{i=1}^{k} f(R_i,\vec{r}_i) \qquad (14)$$

$$\delta E = \sum_{i=1}^{k} \int \delta\varepsilon(R_i,\theta,\vec{r}_i) f(R_i,\vec{r}_i) dR_i d\vec{r}_i = \int \sum_{i=1}^{k} \delta\varepsilon(R_i,\theta,\vec{r}_i) \prod_{i=1}^{k} f(R_i,r_i) dR_i d\vec{r}_i. \qquad (15)$$

Here, $\delta\varepsilon(\vec{r}_1,R_1,\theta) = \delta\varepsilon_1(\vec{r}_1,R_1,\theta) + \delta\varepsilon_{int}(\vec{r}_1,R_1,\theta)$ is the energy of a local configuration of the pore and its environment. According to Eqs. (14) and (15), a change in the energy of the porous medium at the extrusion of the liquid from filled pores can be represented in the form of the sum of the energies of the local configuration of the pore and its environment.

## 3. Energy spectrum and density of states of local configurations

According to Eq. (13), each pore in the percolation cluster of filled pores is in the environment of a certain number of filled and empty pores. Since the percolation cluster is inhomogeneous, the number of pores surrounding any pore belonging to the percolation cluster is individual. Consequently, the percolation cluster of filled pores can also be considered as a set of its pores together with the local environments (configurations) of them. Therefore, as follows from Eqs. (10), (11), (13), and (15), the state of the filled porous medium is a set of pores together with the local environments (configurations) of them. According to Eqs. (10) and (13), these local configurations are connected with each other through menisci in neighboring pores and through the «external field» determined by the degree of filling $\theta$, which appears because of the existence of the percolation cluster. Each local configuration at the degree of filling $\theta$ corresponds to the presence of filled pores belonging to the percolation cluster and empty pores (10) around the filled pore with the radius $R$.

Using Eqs. (11) and (13), the energy $\delta\varepsilon$ of the local configuration containing $n$ empty pores can be represented in the form [12, 13]

$$\delta\varepsilon(\theta,R) = \frac{1}{z}\sum_{n=0}^{z-1} \delta\varepsilon_n,$$

$$\delta\varepsilon_n = -4\pi R^2 \delta\sigma(1-q(\frac{R_0}{R})^\alpha) + 4\pi q R^{2-\alpha} R_0^\alpha \sigma(1-\theta)^n (P(\theta))^{z-n} \frac{z-2n}{z} \frac{z!}{n!(z-n)!}. \quad (16)$$

Here, $P(\theta)$ is the probability for the filled pore to belong to the percolation cluster, $R_0 \sim \vec{R}/z$ is the minimum radius in the pore size distribution, $\alpha \approx 0.3$, and $q \sim 1$ [12, 13]. It is seen that the second term in Eq. (16) is related to the change in the number of menisci at the extrusion of the liquid from pores and is the energy of the «multiparticle interaction» between liquid clusters in pores of the local configuration in the $n$ th state. It can be both positive at $n < z/2$ and negative at $n > z/2$. As a result, $\delta\varepsilon_n$ at a given degree of filling $\theta_1$ can be both positive and negative, depending on the number $n$. A negative $\delta\varepsilon_n$ value means that the extrusion of the liquid from pores in the $n$ th state is energetically favorable. A positive $\delta\varepsilon_n$ value means the existence of the energy barrier for the decay of the metastable configuration in this state. Thus, Eq. (16) indicates the existence of local maxima and minima of the energy necessary for the extrusion of the liquid from the $n$ th state. We now calculate the density of states $g(E,\theta)$ corresponding to the presence of the local configuration in a state with the energy $E$ at the degree of filling $\theta$:

$$g(E,\theta) = \int_0^\infty dR \sum_{n=0}^{z-1} \delta(E - \delta\varepsilon_n(R)) f(R). \quad (17)$$

Figure 1 shows the energy dependence of the density of states of the local configuration for various degrees of filling $\theta$. The calculations were performed with the results for the surface tension coefficient $\sigma(T)$ from [38]. The surface tension coefficient at $T = 293$ K is 75 mJ/m$^2$ [43]. The $\delta\sigma$ value at $T = 293$ K is 22 mJ/м$^2$. The quantity $R_0 \sim \frac{\overline{R}}{z}$ was estimated within the model of randomly distrbuted spheres [41, 42].

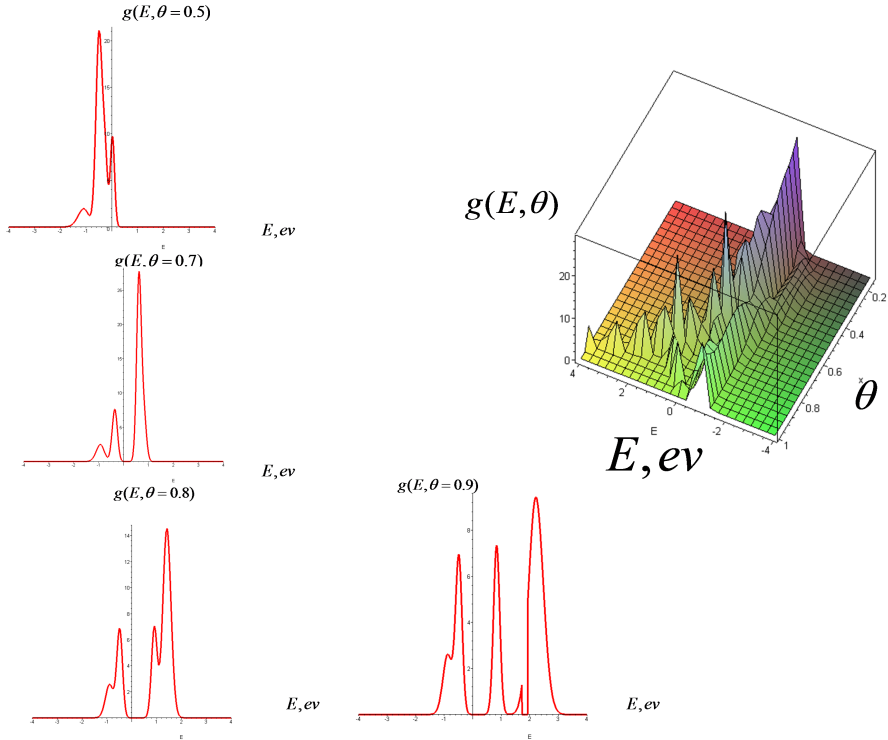

Fig. 1. Density of states of the local configuration versus the energy and degree of filling of the medium $\theta$.

It is seen in Fig. 1 that the states of the system at the degree of filling $\theta = 0.5$ are concentrated at negative energies of local configurations $E < 0$ or at energies below the temperature $E < T$.

The characteristic liquid extrusion time $\tau$ for these states is determined by the hydrodynamic time $\tau_0$ of the motion of the liquid in the porous medium. At the degree of filling $\theta = 0.7$, some local configurations are in states with a negative energy. The liquid flows out of these configurations in the hydrodynamic time. The other part of the liquid is in a state for which the potential barrier is positive and amounts $E \sim 1\,ev$. The characteristic liquid extrusion time for these states is determined by the potential barrier for extrusion and can significantly exceed the hydrodynamic time, $\tau \gg \tau_0$, when the barrier is above the temperature, i.e., $\delta A > T$. These configurations correspond to a metastable state of the liquid with the decay time determined by the barrier. At the degree of filling $\theta = 0.8$, the number of local configurations with a negative energy $E < 0$ decreases, whereas the number of local configurations with a positive barrier increases. At the same time, states with a higher barrier, $E > 1.5\,ev$, appear in addition to states with $E \sim 1\,ev$. These configurations correspond to the appearance of additional metastable states of the liquid with a longer decay time, which is determined by the height of the barrier.

## 4. Mechanism of Slow Relaxation of the system of strongly interacting liquid clusters in the disordered nanoporous medium according to the self-organized criticality scenario

From the analysis made above follows that the extrusion of the liquid from different local configurations of pores can take different times, and the number of configurations of pores involved in extrusion depends on the observation time. As a result, the state of the system relaxes and the amount of the liquid remaining in the porous medium $\theta(t)$ depends on the observation time.

To calculate this dependence following [12, 13], we introduce the distribution function $F^V(t)$ for the extrusion times from a pore with the radius $R$ and volume $V = (4\pi/3)R^3$. This function specifies the volume fraction $d\theta(t) = F^V(t)dt$ of pores from which the liquid is extruded in the time $dt$. For a nonrandom liquid extrusion time $\tau$, $F^V(t) = \delta(t-\tau)$, where $\delta(t)$ is the Dirac delta function.

The relaxation function $F^V(t)$ for the state $\theta(t)$ is normalized to unity, $\int_0^\infty d\tau F^V(\tau) = 1$, and the integral of $F^V(t)$ gives the volume fraction of pores $\theta(t)$ from which the liquid is extruded in the time interval from zero to $t$:

$$\int_0^t d\tau F^V(\tau) = \theta(t). \tag{18}$$

Then, if the medium at the initial time (time of the formation of the metastable state) is filled to the degree of filling $\theta(0) = \theta$, the volume fraction $\theta(t)$ of pores in which the liquid remains to the time $t$ is given by the expression

$$\theta(t) = \theta \int_t^\infty dt F^V(t).$$

It follows from the definition of $g(E)$ that

$$F^V(t) = \int_{-\infty}^\infty \delta(t-\tau(E))g(E)dE = \int_0^\infty \sum_{n=0}^{z-1} \delta(t-\tau_n(R))f^V(R)dR = \sum_{n=0}^{z-1} F^V{}_n(t)$$

$$F_n(t) = \int_0^\infty \delta(t-\tau_n(R))f^V(R)dR.$$

(19)

Here, $\tau_n$ is the relaxation time of the $n$th metastable configuration with the radius $R$ at the degree of filling $\theta$ and $f^V(R)$ is the pore volume distribution function.

As follows from Eqs. (19), the extrusion time distribution function is represented in the form of the sum of partial distribution functions corresponding to extrusion from the local configuration in the $n$th state with the relaxation time $\tau_n$. According to [43], the probability of a change in the state of the system in unit time at the extrusion of the liquid from the $n$th local configuration under the action of fluctuations in the system can be written in the form $w_n = w_0 exp(-\delta\varepsilon_n(\theta,R)/T)$ where $w_0$ is the pre-exponential factor reflecting the dynamics of the extrusion of the liquid from the porous medium. Then, the relaxation time is

$$\tau_n = \tau_0 exp(\delta\varepsilon_n(R,\theta,T)/T) . \quad (20)$$

It follows from Eq. (20) that the relaxation time $\tau_n$ of the $n$th configuration is determined by the waiting time $\sim exp(\delta\varepsilon_n(R,\theta,T)/T)$ necessary for overcoming the local barrier $\delta\varepsilon_n(R,\theta,T)$ and by the time $\tau_0$ of the subsequent barrierless hydrodynamic extrusion of the liquid from the local configuration. According to Eq. (20), the quantity $\delta\varepsilon_n(R,\theta,T)$ serving as the barrier for the $n$th local configuration depends on the radius of the pores, the probability of belonging to the percolation cluster $P(\theta)$, and the degree of filling $\theta$ of the porous medium with the liquid and, thereby, according to Eq. (16), it depends on the interaction between local configurations in the percolation cluster. The decay of the $n$th metastable state is accompanied by a decrease in the number of filled pores in the percolation cluster and, hence, in the number of neighboring pores. Therefore, the energy of the multiparticle interaction in Eq. (16), as well as thereby the lifetime of local configurations with different numbers $n$, decreases. As a result, the decay of the $n$th local metastable configuration reduces the decay time of local metastable configurations with different numbers $n$ and with different, initially higher energy. In this case, the dependence of the interaction between local configurations on the number of filled pores belonging to the infinite percolation cluster of filled pores serves as an internal feedback at the decay of metastable states in a wide region of the degree of filling $\theta > \theta_c$, where $\theta_c$ is the percolation threshold.

As follows from Eq. (19), the volume fraction $\theta(t)$ of pores from which the liquid is extruded in the time interval from zero to $t$ can be represented in the form

$$\theta(t) = \sum_{n=0}^{z-1} \theta_n(t),$$
$$\theta_n(t) = \theta \int_0^t dt \int_0^\infty \delta(t - \tau_n(R)) f(R) dR. \quad (21)$$

According to Eqs. (21), the amount of the liquid $\theta(t)$ remaining in the porous medium is the sum of time variations of the volume fraction $\theta_n(t)$ of filled pores in the local configuration in the state $n < z-1$. According to Eq. (20), each $n$th state in local configurations relaxes in the time determined by the energy barrier $\delta\varepsilon_n(R,\theta,T)$ given by Eq. (16). According to Eq. (16), $\delta\varepsilon_n$ can be both positive and negative.

We now discuss the relaxation of the initial state of the system corresponding to complete filling $\theta = 1$. The first stage involves the decay of weakly bound states for which $\delta\varepsilon_n < T$. At this stage, the liquid is removed from weakly bound states of filled pores in the hydrodynamic time $\tau_0 \sim 0.1 сек$ and, therefore, the volume fraction of the liquid in pores $\theta_n(t)$ given by Eqs. (21) in these states decreases at times $\sim \tau_0$. This stage results in the formation of the metastable state of filled pores in the porous medium. This metastable state corresponds to bound states of interacting local configurations of filled pores in the percolation cluster for which the energy barrier is positive $\delta\varepsilon_n > 0$, because the energy of the effective «multiparticle attraction» given by Eq. (10) is positive, $\delta\varepsilon_{int} > 0$, and $\delta\varepsilon_{int} > |\delta\varepsilon_1|$. Such states relax at times $t \sim t_n \sim \tau_0 \exp(\delta\varepsilon_n/T) \gg \tau_0$. The formed metastable state decays in times $t > t_0 \sim \max t_n$. The decay through the discussed mechanism ends at the percolation threshold $\theta \approx \theta_{\tilde{n}}$ when the percolation cluster of filled pores disappears. Further relaxation can occur through the mechanism of evaporation from individual filled pores [15]. According to Eq. (21), the volume fraction $\theta(t)$ of pores from which the liquid is extruded in the time interval from zero to $t$ can be represented in the form of the sum of terms corresponding to various relaxation times $\tau_n$. The dependence $\theta_n(t)$ at times $\tau_0 \ll t$ can be calculated under the assumption that most of filled pores with the pore size distribution $f^V(R)$ in the local configuration in the $n$th state at these times are connected through the percolation cluster of filled pores [15]. The calculations similar to [15] gives

$$\theta_n(t) \sim \theta_{np}(\frac{\tau_{qn}}{t})^{a_n},$$

$$a_n = \frac{1}{1+(2-\alpha)\frac{\Delta R}{\bar{R}}\frac{\varepsilon_{n\max}}{T}}, \quad (22)$$

$$\tau_{qn} \sim \tau_0 \exp(\frac{\varepsilon_{n\max}}{T}),$$

$$\theta_{np} \sim \theta \int_0^\infty \eta(\delta\varepsilon_n(R))dR f^V(R).$$

where $\eta(X)$ is the Heaviside step function and $\varepsilon_{n\max} = \max \delta\varepsilon_n$ is the maximum energy barrier for the decay of the local metastable configuration in the $n$th state.

As follows from Eqs. (22), the exponents $a_n$ and times $\tau_{nq}$, which enter into Eq. (23) and specify the relaxation law of the state of the $n$th local configuration, are determined by the energy barrier $\varepsilon_{n\max}$. The volume fraction of the liquid remaining in the local metastable configuration in the $n$th state decreases according to a power law with the exponent $a_n$ and charateristic time $\tau_{nq} \sim \tau_0 \exp(\frac{\varepsilon_{n\max}}{T})$. According to Eqs. (22), different states $\theta_n(t)$ relax with different charateristic times $\tau_{nq}$ and different exponents $a_n$. The slowest relaxing state of the local configuration corresponds to $\tau_{nq} = \max$ and $a_n = \min$ and, hence, $\varepsilon_{n\max} = \max$; whereas the slowest relaxing state to $\tau_{nq} = \min$ and $a_n = \max$ and, therefore, $\varepsilon_{n\max} = \min$.

It follows from Eq. (21) that the decay of the local metastable configuration in the $n$th state results in the disappearance of the corresponding term in sum (21) and, consequently, in a decrease in the degree of filling $\theta$. As a result, according to Eq. (16), the energy of the interaction between local configurations decreases and, therefore, the energy barrier $\delta\varepsilon_m(\theta)$ for extrusion from other states changes. Hence, the exponents $a_n(\theta(t))$ and volume fractions $\theta_{np}(\theta(t))$, which are given in Eqs. (22) and determine the relaxation time $\tau_{qn}(\theta(t))$ of the state, depend on the current degree of filling $\theta(t)$. For this reason, according to Eqs. (21) and (22), the volume fraction of the liquid remaining in the metastable state in the porous medium should be determined from the self-consistent equation

$$\theta(t) = \sum_{n=0}^{z-1} \theta_{np}(\theta(t))(\frac{\tau_{qn}(\theta(t))}{t})^{a_n(\theta(t))}. \quad (23)$$

To analyze this equation, we note that the terms in sum (23) include different exponents $a_n(\theta(t))$ and times $\tau_{qn}(\theta(t))$ according to Eqs. (22). As a result, the relaxation of the system determined by the dependence $\theta(t)$ at each time instant occurs as the «rapid» relaxation of the $n$th state, which is determined by the maximum exponent $a_n$ and the minimum time $\tau_{qn}$, and as the «slow» relaxation of the $n$th current state, which is determined by the minimum exponent $a_n$ and the maximum time $\tau_{qn}$. As follows from Eq. (23), the «rapid» relaxation of the $n$th local metastable state occurs through extrusion from the local configuration of pores with the minimum energy barrier $\varepsilon_{n\max}(\theta)$ (16), which is determined by the interaction between local configurations. According to Eq. (23), the relaxation of the entire system, which is determined by the dependence $\theta(t)$, at each time instant occurs through the relaxation of the $n$th current state determined by the slowest relaxing «mode» with the minimum exponent $a_n$ and the maximum time $\tau_{qn}$. It follows from Eq. (21) that the decay of the local metastable configuration in the $n$th states with the energy barrier $\delta\varepsilon_n$ leads to a decrease in the degree of filling $\theta$ and, consequently, according to Eq. (16), to a decrease in the energy barrier and extrusion time (20) for other local metastable configurations that were initially in states with the energy barrier $\delta\varepsilon_m > \delta\varepsilon_n$. Thus, according to Eq. (23), the relaxation of the metastable state at times $t \gg \tau_0$, $t \sim t_0$ occurs through the «rapid» decay of «short-lived» states (modes), which «are adjusted» to the current degree of filling $\theta(t)$ because of the dependence of $\theta_{iд}$, $\tau_{qn}$ (22), ensuring feedback with the «slow» relaxation of the $n$th state determined by the minimum exponent $a_n$ and the maximum time $\tau_{qn}$. According to Eq. (23), the set of such modes contributes to the total relaxation of the metastable state at times $t \sim t_0 \gg \tau_0$. Therefore, the relaxation of the metastable state in times $\tau_0 \ll t \sim t_0$ corresponds to «the trajectory of motion» of the system in the ($\theta, n$) space with the slowest relaxation of the local metastable configurations in the $n$th states. The relaxation of the state of the local metastable configurations on this trajectory occurs because of feedback with the minimum exponent $a_n$ and maximum time $\tau_{qn}$. The extrusion of the liquid from the local configuration in this state is accompanied by the extrusion of the liquid from neighboring configurations with close heights of the energy barrier $\delta\varepsilon_n$, resulting in avalanche extrusion from such configurations. For this reason, the state on the «trajectory of motion» of the system in the ($\theta, n$) space with the slowest relaxation of local metastable configurations can be considered as critical. Figure 2 shows the energy of the state of the local configuration $\delta\varepsilon_n$ corresponding to the $n$th mode for the configuration of the pore with the radius equal to the average radius $R = \overline{R}$ for a Gaussian pore volume distribution $f^V(R)$. The calculations were performed with the same parameters as for Fig. 1.

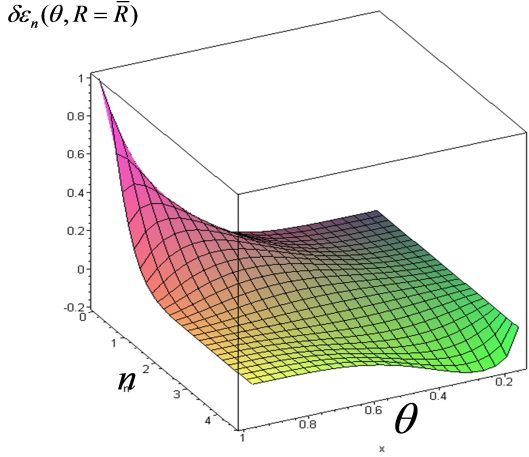

*Fig. 2. Energy of the state of the local configuration $\delta\varepsilon_n$ corresponding to the nth mode for the configuration of the pore with the radius equal to the average radius $R = \bar{R}$ for a Gaussian pore volume distribution $f^V(R)$. The states on the crest correspond to the critical state of the system.*

It is seen in Fig. 2 that the energy of the state of the local configuration $\delta\varepsilon_n$ as a function of the variables $\theta, n$ has a maximum (crest) corresponding to the states and degrees of filling of local metastable configuration with the slowest relaxation, according to Eqs. (20) and (22). Consequently, the states on the crest determine the critical state of the system throughout the range $\theta > \theta_c$. According to Eq. (23), the time dependence of the degree of filling for the critical state is given by the expression

$$\theta(t) \approx \theta_{n(\theta(t))p}(\theta(t))(\frac{\tau_{qn(\theta(t))}(\theta(t))}{t})^{a_{n(\theta(t))}(\theta(t))}. \quad (24)$$

Here, $n(\theta(t))$ is the number corresponding to the state of the local configuration on the slowest relaxing trajectory (critical state).

As follows from Eqs. (22), $a_{n(\theta)} \approx const \approx a_0$ and $\tau_{n(\theta)q} \approx const \approx \tau_{n=0} = \tau_q$ for the slowest relaxing mode in view of the slow power-law dependence $\theta(t)$. Therefore, for the trajectory corresponding to the slowest relaxation, it follows from Eq. (24) that

$$\theta(t) \approx \theta_p(\frac{\tau_q}{t})^a, \quad (25)$$

$$a = \frac{1}{1+(2-\alpha)\frac{\Delta R}{\bar{R}}\frac{\varepsilon_z}{T}},$$

$$\tau_q \sim \tau_0 \exp(\frac{\varepsilon_z}{T}),$$

$$\theta_p \sim \theta \int_0^\infty \eta(\delta\varepsilon_{n=0}(R))dR f^V(R).$$

where $\varepsilon_{ji}$ is the highest energy barrier for the decay of the state of the local metastable configuration in the critical state on the trajectory $n(\theta(t))$.

Expressions (25) describe the relaxation of the formed metastable state corresponding to the mean-field picture considered in [15], where the formed metastable state corresponding to bound states of interacting local configurations of filled pores relaxes slowly according to a power law with the exponent $a$ given in Eqs. (24) in times $t \sim \tau_q$. The formed metastable state decays at times $t > t_0$ because of a decrease in the degree of filling $\theta$ and, as a result, a decrease in the energy barrier for extrusion $\varepsilon(R,\theta_1)$. In this case, the $\varepsilon_z$ value decreases and the exponent $a$ increases according to Eqs. (25). Consequently, the decay rate of the metastable state increases with the relaxation of the system.

The final expressions indicate that the relaxation of the system occurs through the decay of the local metastable configuration responsible for the formation of critical conditions for the decay of other local metastable configurations with a different, initially higher energy. The decay of the metastable state of some local configurations reduces the number of filled pores in this percolation cluster and, therefore, the number of neighboring pores. As a result, the energy of the multiparticle interaction decreases and critical conditions appear for the decay of other metastable configurations. In this case, the system relaxes on the trajectory in the (configuration number, degree of filling) space corresponding to the critical state with the slowest relaxing mode. The time dependence of the relative volume $\theta$ of the confined liquid on this trajectory has the form of a power law given by Eqs. (25). It corresponds to a relaxation picture in approach of an average field [15].

To summarize, the relaxation of the system of the nanoporous medium with the nonwetting liquid is a self-organized criticality process characterized by waiting for fluctuation necessary for overcoming a local metastable energy state with the subsequent avalanche decay of local metastable configurations of pores. The dependence of the interaction between local configurations on the number of filled pores belonging to the infinite percolation cluster of filled pores serves as an internal feedback initiating the SOC process.

The model of the relaxation of the porous medium with the nonwetting liquid developed in this work demonstrates possible mechanisms and scenarios of SOC for disordered atomic systems.